# Non-left-handed transmission and bianisotropic effect in a $\pi$-shaped metallic metamaterial

Zheng-Gao Dong,[1,*] Shuang-Ying Lei,[2] Qi Li,[1] Ming-Xiang Xu,[1] Hui Liu,[3] Tao Li,[3] Fu-Ming Wang,[3] and Shi-Ning Zhu[3]

[1]*Physics Department, Southeast University, Nanjing 211089, People's Republic of China*
[2]*Key Laboratory of MEMS of Ministry of Education, Southeast University, Nanjing 210096, People's Republic of China*
[3]*National Laboratory of Solid State Microstructures, Nanjing University, Nanjing 210093, People's Republic of China*

A $\pi$-shaped metallic metamaterial (geometrically, a combination medium of C-shaped resonators and continuous wires) is proposed to investigate numerically its transmission band near the resonant frequency, where otherwise should be a negative-permeability (or negative-permittivity) stop band if either the C-shaped or continuous-wire constituent is separately considered. However, in contrast to the left-handed materials (LHM) composed of split ring resonators and wires as well as other metallic LHM, this resonant transmission is a non-left-handed one, as a result of the intrinsic bianisotropic effect attributed to the electrically asymmetric configuration of this $\pi$-shaped metamaterial.

# I. INTRODUCTION

Metamaterials and left-handed materials (LHM) are now amongst the most popular artificial materials for their striking abilities of extending the application range of previous electromagnetic media.[1-3] As is well known, metamaterials are generally composed of shaped elements with periodic intervals much smaller than the incident electromagnetic wavelengths, and thus they can be referred to as effective homogeneous media with electromagnetic constitutive parameters (i.e., electric permittivity $\varepsilon$ and magnetic permeability $\mu$), which are usually unavailable from naturally occurring materials. For example, a spilt-ring metallic metamaterial extends magnetic response up to terahertz regime, and even visible frequencies;[4,5] while a continuous-wire metallic metamaterial keeps negative $\varepsilon_{\text{eff}}$ at frequencies as low as desired.[6]

As a kind of metamaterial, LHM require simultaneously negative electric and magnetic responses (i.e., $\varepsilon_{\text{eff}} < 0$ and $\mu_{\text{eff}} < 0$), and thus are more difficult to be designed than other metamaterials with only electric or magnetic response. Experimentally, the most intensively studied LHM are generally composed of metallic elements because of their extraordinary $\varepsilon$ under the plasma frequencies. Up to date, several metallic metamaterials, such as the $\Omega$-shaped,[7] S-shaped,[8] H-shaped,[9] and double-wire media,[10-12] have been claimed to be LHM, in addition to the firstly invented one, that is, a composite metamaterial with metallic split ring resonators (SRR) and continuous wires.[3] However, SRR-wire metamaterials as LHM are not widely accepted as soon as reported because of the difficulties in interpreting its resonant transmission. Meanwhile, Zhou *et al.* believed that it was not easy to obtain LHM with only pairs of cut wires unless they were replaced by a composite medium of continuous and cut wires,[10,13] while conversely the negative index of refraction was reported to be existed in a metamaterial with pairing gold rods (i.e., cut wires).[11] Furthermore, some feasible criteria were proposed in order to distinguish between the left-handed and right-handed transmission peaks.[14-16] All this investigations suggest that it is not always the case for a resonant transmission band of metallic metamaterial to be left-handed, and thus it needs rather careful before making an unambiguous conclusion of reaching a true LHM. In this paper, a $\pi$-shaped metallic metamaterial is proposed to investigate numerically its resonant transmission by comparing the similar to and difference from the left-handed transmission of SRR-wire metamaterial. It is found that such a $\pi$-shaped metallic metamaterial does not result in LHM because of the underlying bianisotropic effect.

## II. STRUCTURAL MODEL

Figures 1(a) and 1(b) show a structured metallic array and the corresponding physical unit cell with geometric scales, respectively. This medium is called as $\pi$-shaped metamaterial for convenience [as outlined by the rectangular dashed line in Fig. 1(a)]. This $\pi$-shaped metamaterial, just in terms of geometry, can be conceptually referred to as a combination of metallic C-shaped resonators and continuous wires [Figs. 1(c) and 1(d), respectively]. As is commonly reported, metamaterials with spirally conducting path for induced current oscillation are the general way to get magnetism above gigahertz frequencies,[4,5,12,17,18] while a continuous-wire metamaterial exhibits negative $\varepsilon_{\text{eff}}$ under its plasma frequency.[6]

The interest in such a $\pi$-shaped metamaterial is to investigate whether or not there exists left-handed electromagnetic response in certain frequency band since this medium is designed as a combination of metallic C-shaped spirals and continuous wires, as an analogy to the other metallic LHM. To excite the negative magnetic response of its C-resonator constituent as well as the negative electric response of its continuous-wire constituent, the incident electromagnetic waves are polarized with magnetic field perpendicular to the C-resonator plane ($\vec{H} // x$-axis) and electric field parallel to the $y$-axis. Consequently, the propagation direction (wave vector $\vec{k}$) is along the $z$-axis. In our numerical simulations based on the full-wave finite element method, the polarized incidence condition is satisfied throughout this paper by applying corresponding perfect magnetic and electric boundaries (corresponding to the boundaries perpendicular to the $x$- and $y$-axes, respectively), and the waves are incident from one of the two wave ports (i.e., the two boundaries perpendicular to the $z$-axis). For simplicity, all the simulations are performed in the background of vacuum without taking the substrate into account. Additionally, the influence of metal losses on the transmission characteristics is another important issue of metallic metamaterials. To decrease the loss factor to the best and without losing generality, the metallic elements under simulations are assumed to be "perfect" conductor (with the parameters of $\sigma = 1.0 \times 10^{30}$ Siemens/m and $\varepsilon = \mu = 1$), which can be approximately realized by metals in the microwave frequencies.

### III. RESULTS AND DISCUSSION

## A. the similar to LHM

As is known, the SRR-wire composite metamaterial was originally found to be LHM according to the phenomenon that its passband occurred just at the stop bands of both SRR-only medium ($\mu_{eff}$ <0) and continuous-wire medium ($\varepsilon_{eff}$ <0).[19,20] Accordingly, in this paper, the transmission coefficients (i.e., the scattering parameter $S_{21}$) of both the $\pi$-shaped metamaterial and its individual constituents, namely, the C-resonator-only (with parameter $l$ =0) and continuous-wire metamaterials, are investigated together for comparison. As far as the transmission spectrum of the C-resonator-only medium is concerned (Fig. 2, open square line), there is a stop band located in the range from 8.4 to 12 GHz, expectedly resulted from $\mu_{eff}$ <0. A retrieval procedure[21-23] based on the scattering parameters of $S_{21}$ and $S_{11}$ is furtherly developed to confirm its negative magnetic response. As is presented by the solid squares in Fig. 2, the retrieved result indicates that there is a $\mu_{eff}$ <0 regime corresponding to the resonant stop band. Note that the stop band for the C-shaped configuration under study is broader than an SRR usually exhibits, which is due to the fact of electric coupling to the magnetic resonance.[16] As for the transmission characteristic of a continuous-wire medium, the results in Fig. 3 show that no transmission below the electric plasma frequency of about 20 GHz is due to negative $\varepsilon_{eff}$, in consistent with literatures.[6,21]

On the other hand, the transmission spectrum of a $\pi$-shaped metamaterial shows, intriguingly, an obvious passband around 11 GHz (Fig. 4), while this frequency regime, as was specified before, shows no transmissions for not only the C-resonator-only (Fig. 2), but also the continuous-wire metamaterials (Fig. 3). In another way, a gap-closing medium [see Fig. 1(b), but with $g = 0$] is investigated for further clarity, and the result exhibits that a gap-closing medium is essentially a metamaterial with negative $\varepsilon_{\text{eff}}$ since there is no qualitative difference to a continuous-wire medium in terms of both their transmission spectra and retrievals. Note that the gap-closing method[14-16] was introduced as an easy-to-apply criterion to identify the left-handed transmission of SRR-wire metamaterials. Consequently, the question is whether or not it is a transmission band under the principle of left-handed electromagnetic mechanism.

### B. the difference from LHM

Firstly, to make a further investigation on the resonant transmission of $\pi$-shaped metamaterial, the width of the continuous-wire constituent $w$ [represented in the inset of Fig. 5(a)] is tuned from 0.1 mm to 1.2 mm, while the width of its C-shaped branch is kept unchanged. Fig. 5(a) reveals interestingly that the resonant transmission around 11 GHz is more and more suppressed with the width $w$. By contrast, as for the SRR-wire metamaterials, the resonant left-handed transmissions were never suppressed when the originally used thin wires were replaced by widened wires or by more wires in number.[24-26] Fig. 5(b) reduplicates it by modifying only the width $w'$ of continuous wires, with the other simulation conditions identical to Ref. 26.

Secondly, Fig. 5(a) also suggests that the transmission peaks are located at the left side to the resonant frequencies, while for the SRR-wire LHM, the corresponding peaks are found at the right side to the resonant frequencies [Fig. 5(b)]. This discrepancy should be important since generally the constitutive parameters are only negative above and next to the resonant frequencies (i.e., right side).

Thirdly, we recently reported that the negative refraction phenomenon of an SRR-wire LHM can be confirmed by simulating the wedge-shaped configuration constructed by plenty of SRR and wires.[26] This method is also adopted to investigate in what manner the transmitted waves around 11 GHz are bent by the $\pi$-shaped metamaterial in wedge-shaped configuration. In our simulations, no transmitted beam is found to be negatively refracted with backward phase direction.

Additionally, it is worth mentioning that all of the above results are kept qualitatively even if the scales in Fig. 1(b) are changed.

### C. Bianisotropic effect and retrievals

What has been investigated before is that the resonant transmission of $\pi$-shaped metamaterial is different from that of SRR-wire LHM in some aspects, while in other ways it seems as if they were the same. In fact, the ambiguity would be solved if the attention is drawn from the resonant transmission to the $\pi$-shaped configuration itself, and taking into account that such a $\pi$-shaped metamaterial is an electrically asymmetric structure, and thus intrinsically a bianisotropic medium. Generally speaking, a bianisotropic medium means that its electric and magnetic responses are coupled to each other,[27-29] and hence an additional parameter (namely, the electromagnetic coupling coefficient $\xi$) is required to modify the general constitutive equations. As a matter of fact, the relation between bianisotropic materials and LHM, to the best of our knowledge, is yet to be systematically established because the additional $\xi$ is not considered at all in the LHM system. Consequently, it is hard to conclude that the resonant transmission from $\pi$-shaped metamaterial is a left-handed response.

Nevertheless, in another way, assume the $\pi$-shaped medium is not strong enough, but negligible, in its bianisotropic effect such that the generally used retrieval procedure[21,22] can be adopted to find what will happen to its effective constitutive parameters with respect to the resonance. Figs. 6(a-d) are the retrieved results carried out to determine the effective impedance $z_{\text{eff}}$, index of refraction $n_{\text{eff}}$, and then $\varepsilon_{\text{eff}}$ and $\mu_{\text{eff}}$ of $\pi$-shaped metamaterial. Surprisingly, the retrieved result obtained according to Refs. 21 and 22 is physically reasonable. It reveals that neither negative $n_{\text{eff}}$ nor simultaneously negative $\varepsilon_{\text{eff}}$ and $\mu_{\text{eff}}$ are obtained around 11 GHz. Instead, the resonance causes a *positive* $\varepsilon_{\text{eff}}$ band around the resonance (which otherwise should be negative) and keeps the positive sign of $\mu_{\text{eff}}$ unchanged. This result can be interpreted in a reasonable way. Consider that the electrically induced current in a $\pi$-shaped medium is rather overwhelming than the magnetically induced one, such that it destroys completely the possibility of negative magnetic response which otherwise would be resonantly existed relying on the C-shaped branch of $\pi$-shaped metamaterial. As a result, no matter whether the bianisotropic effect is involved in or not, the resonant transmission band around 11 GHz should be a non-left-handed electromagnetic response, in contrast to the possible mistake just conjectured by comparing the passbands and stop bands of several geometrically relative media.

It is emphasized that the objective in this paper is to demonstrate an example of non-left-handed resonant transmission (despite that in some ways it seems as if it were left-handed), which is reported in seldom by the published papers. Therefore it is not a necessary issue to resort to the rather complicated retrieval procedure in Ref. 30 for solving the electromagnetic coupling coefficient $\xi$. By the way, it is worth mentioning that the $\pi$-shaped medium under study can be used in microwave application as a frequency selective surface, for its controllable transmission characteristic at the narrow band around 11 GHz.

## IV. SUMMARY

In summary, a $\pi$-shaped metallic metamaterial has been numerically investigated to study its resonant transmission band through scattering parameters simulations, retrievals, as well as analysis based on the knowledge of bianisotropic effect. The results demonstrate that, for a metallic metamaterial with spirals (e.g., C-shaped resonators) and continuous wires, it is not always the case for its resonant transmission with simultaneously magnetic and electric responses to be left-handed, especially when bianisotropic effect is inherent in the geometric configuration. It is expected that the conclusion in this investigation would be helpful to understand the ambiguity on left-handed and/or non-left-handed transmissions in various resonant metallic metamaterials.

## ACKNOWLEDGMENTS

This work was supported by the National Natural Science Foundation of China under contract No. 10604029, and the Natural Science Foundation of Jiangsu

Province (No. BK2006106).

[*]Corresponding author, Electronic address: zgdong@seu.edu.cn

[1]D. R. Smith, J. B. Pendry, and M. C. K. Wiltshire, Science **305**, 788 (2004).

[2]J. B. Pendry and D. R. Smith, Phys. Today **57**, 37 (2004).

[3]R. A. Shelby, D. R. Smith, and S. Schultz, Science **292**, 77 (2001).

[4]T. J. Yen, W. J. Padilla, N. Fang, D. C. Vier, D. R. Smith, J. B. Pendry, D. N. Basov, and X. Zhang, Science **303**, 1494 (2004).

[5]C. Enkrich, M. Wegener, S. Linden, S. Burger, L. Zschiedrich, F. Schmidt, J. F. Zhou, Th. Koschny, and C. M. Soukoulis, Phys. Rev. Lett. **95**, 203901 (2005).

[6]J. B. Pendry, A. J. Holden, W. J. Stewart, and I. Youngs, Phys. Rev. Lett. **76**, 4773 (1996).

[7]J. Huangfu, L. Ran, H. Chen, X.-M. Zhang, K. Chen, T. M. Grzegorczyk, and J. A. Kong, Appl. Phys. Lett. **84**, 1537 (2004).

[8]H. Chen, L. Ran, J. Huangfu, X. Zhang, K. Chen, T. M. Grzegorczyk, and J. A. Kong, Phys. Rev. E **70**, 057605 (2004).

[9]J. Zhou, T. Koschny, L. Zhang, G. tuttle, and C. M. Soukoulis, Appl. Phys. Lett. **88**, 221103 (2006).

[10]J. Zhou, L. Zhang, G. Tuttle, T. Koschny, and C. M. soukoulis, Phys. Rev. B **73**, 041101(R) (2006).

[11]V. M. Shalaev, W. Cai, U. K. Chettiar, H.-K. Yuan, A. K. Sarychev, V. P. Drachev, and A. V. Kildishev, Opt. Lett. **30**, 3356 (2005).

[12]K. Guven, M. D. Caliskan, and E. Ozbay, Opt. Express **14**, 8685 (2006).

[13]G. Dolling, C. Enkrich, M. Wegener, J. F. Zhou, C. M. Soukoulis, and S. Linden, Opt. Lett. **30**, 3198 (2005).

[14]N. Katsarakis, T. Koschny, M. Kafesaki, E. N. Economou, E. Ozbay, and C. M. Soukoulis, Phys. Rev. B **70**, 201101(R) (2004).

[15]T. Koschny, M. Kafesaki, E. N. Economou, C. M. Soukoulis, Phys. Rev. Lett. **93**, 107402 (2004).

[16]N. Katsarakis, T. Koschny, M. Kafesaki, E. N. Economou, and C. M. Soukoulis, Appl. Phys. Lett. **84**, 2943 (2004).

[17]J. D. Baena, R. Marques, F. Medina, and J. Martel, Phys. Rev. B **69**, 014402 (2004).

[18]A. K. Sarychev, G. Shvets, and V. M. Shalaev, Phys. Rev. E **73**, 036609 (2006).

[19]D. R. Smith, W. J. Padilla, D. C. Vier, S. C. Nemat-Nasser, and S. Schultz, Phys. Rev. Lett. **84**, 4184 (2000).

[20]R. A. Shelby, D. R. Smith, S. C. Nemat-Nasser, and S. Schultz, Appl. Phys. Lett. **78**, 489 (2001).

[21]D. R. Smith, S. Schultz, P. Markos, and C. M. Soukoulis, Phys. Rev. B **65**, 195104 (2002).

[22]D. R. Smith, D. C. Vier, Th. Koschny, and C. M. Soukoulis, Phys. Rev. E **71**, 036617 (2005).

[23]X. Chen, T. M. Grzegorczyk, B.-I. Wu, J. Pacheco, and J. A. Kong, Phys. Rev. E **70**, 016608 (2004).

[24]M. Gokkavas, K. Guven, I. Bulu, K. Aydin, R. S. Penciu, M. Kafesaki, C. M. Soukoulis, and E. Ozbay, Phys. Rev. B **73**, 193103 (2006).

[25]K. Li, S. J. Mclean, R. B. Greegor, C. G. Parazzoli, and M. H. Tanielian, Appl. Phys. Lett. **82**, 2535 (2003).

[26]Z. G. Dong, S. N. Zhu, H. Liu, J. Zhu, and W. Cao, Phys. Rev. E **72**, 016607 (2005).

[27]J. A. Kong, *Electromagnetic Wave Theory*, 2<sup>nd</sup> ed. (Wiley, New York, 1990).

[28]R. Marques, F. Medina, and R. Rafii-El-Idrissi, Phys. Rev. B **65**, 144440 (2002).

[29]D. R. Smith, J. Gollub, J. J. Mock, W. J. Padilla, and D. Schurig, J. Appl. Phys. **100**, 024507 (2006).

[30]X. Chen, B.-I. Wu, J. A. Kong, and T. M. Grzegorczyk, Phys. Rev. E **71**, 046610 (2005).

FIG. 1. (a) Schematic for the array of a $\pi$-shaped metamaterial. The rectangular shorted line outlines the so-called $\pi$ shape. (b) Unit cell with geometric scales $a$ =12.1 mm, $b$ =5.0 mm, $c$ =4.0 mm, $d$ =2.37 mm, $h$ =3.1 mm, $g$ =0.7 mm, and $l$ =2.0 mm. the metal is 0.6 mm in width and 0.05 mm in thickness. The propagation of the polarized electromagnetic wave is along the *z*-axis, with electric field in the *y*-direction while magnetic field in the *x*-direction. The $\pi$-shaped metamaterial is geometrically a combination of C-shaped resonator (c) and continuous wire (d).

FIG. 2. (Color online) Transmission coefficient (open square line) and retrieved permeability (solid square line) for C-shaped medium.

FIG. 3. (Color online) Transmission coefficient (open square line) and retrieved permittivity (solid square line) for continuous-wire medium. The vertical arrow marks from where the effective permittivity begins to be positive.

FIG. 4. Transmission coefficient for the $\pi$-shaped metamaterials.

FIG. 5. (Color online) Comparison of resonant transmissions by tuning only the width of continuous-wire constituents. The tuning parameter $w$ and $w^{'}$ are depicted in the insets, respectively. (a) Suppressed resonant transmission of $\pi$-shaped metamaterial. (b) Unsuppressed resonant transmission of SRR-wire LHM. Note that the two transmission bands are located in different side of their corresponding resonant frequencies (yellow shaded regions).

FIG. 6. (Color online) Retrieved impedance (a), index (b), permeability (c), and permittivity (d) of the $\pi$-shaped metamaterial (without considering its bianisotropic effect).

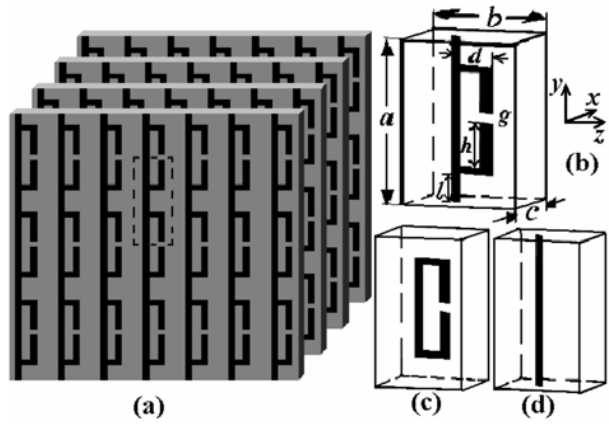

FIG. 2.    Z. G. Dong

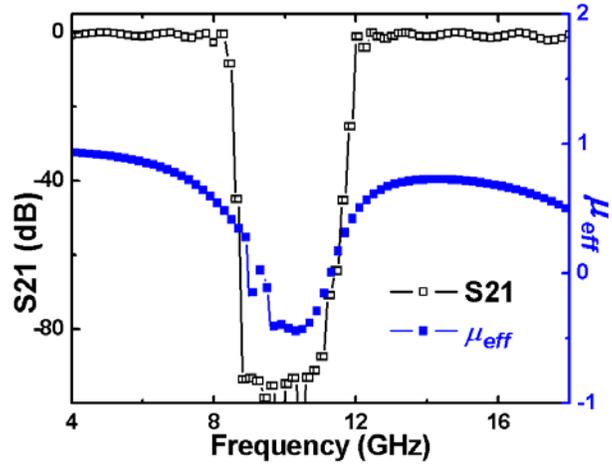

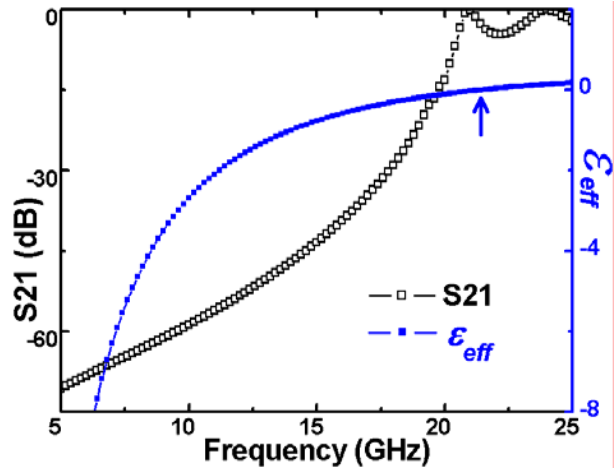

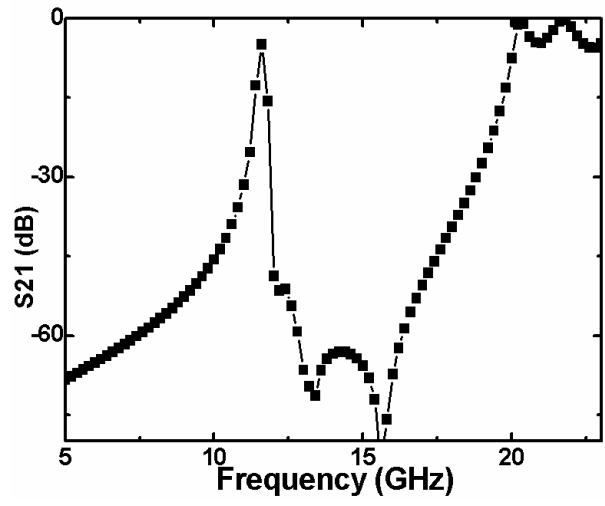

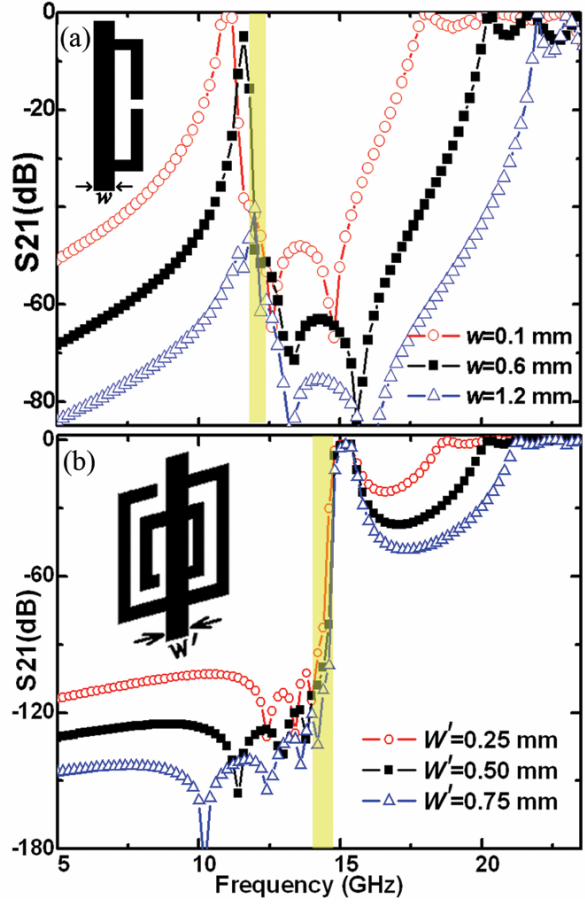

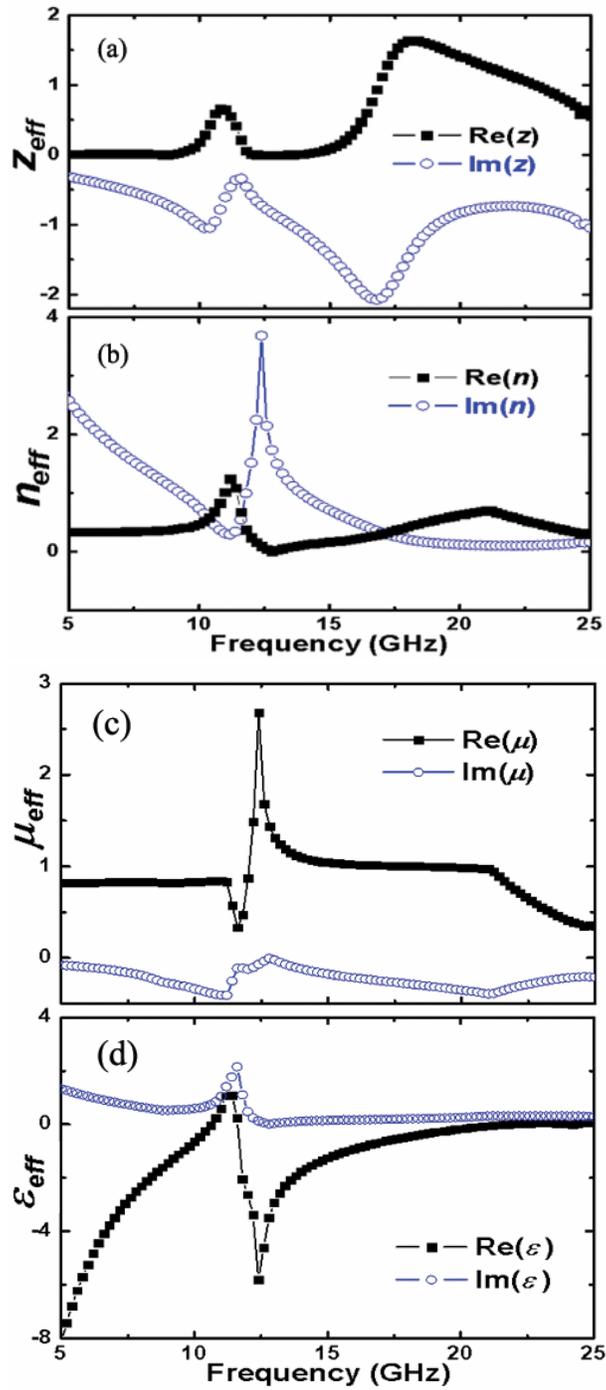